# Open Universe Modeling: Information Layer and Time Dilation

B. Baykant ALAGOZ[*]

**Abstract**: *In this study, we suppose that the universe has an information processing layer, which conveys the informational contents accompanying the physical events. In this manner, universe is considered to be composed of several associative layers such that one rises on the other layer. Preliminary, we present a method for the analytical treatment of the amount of information processed by universe itself, and then we try to show its correspondence with theories developed for the time dilation and gravitational forces.*

---
[*] OncuBilim Algorithm and Systems Labs. E-mail: alagozb@yahoo.com

## 1. Introduction

Throughout the history of empirical sciences, every entity of the physical world and their interaction were seen to have an informational content, which human try to discover. In the journey of the understanding physical world surrounding us, our observations can be said to be an efford to reach the information contained in physical events, states and their interactions. This portrait, solely, can be thought of as evidence for the existence of an information content accompanying our physical world.

More advanced discussions on the informational structure of the physical world have been carried out in many studies: linking fundamental physics with digital computers [1], including the first hypothesis implying the universe is a digital computer [2], the assumption that the universe can be a huge cellular automaton [3] or universal Turing machine [4], more general studies collectively known as "digital physics" suggesting a program for a universal computer which computes the evolution of the universe in real time [5], as well as a quantum computation perceptive of the digital physics [6,10]. A common aspect of these studies is that our universe itself can be a massive, distributed and cellular computational network, and on such a network, the whole physical world can be simulated. Recently, the interest of some of these studies have been turned into finding the cellular automatons or the rules that can output our physical reality. Some remarkable efforts linking issues of space-time and gravity to digital physics were announced.[6,10]

A universal cellular automaton, imitating all physical realities, is a substantial goal of this field. Nowadays, efforts of linking the physical universe to an informational universe are taking more attention due to its potential to contribute to the solution of problems that represent the bottlenecks of modern physics.

According to the general consensus humans are devised to interact with the informational content of the physical world. Our consciousness identifies the universe by a sensory device; situated in our brain. It collects the signals coming from the physical world by sensory organs and processes them to extract the informational content. On this ground, we hypothesize in this study that the informational content is essential, receivable and transmittable, whereas, the so-called physical reality is a perception or a signification of the received informational content coming from the universe we live in. Fashionably, we suppose that the informational content can form the bases on which our perceived physical world can be built on.

Under the light of Information theory [12] and Relativity theory [13], this study was devoted to the discussion of the transmission properties of the information processing network that carries the informational content of the universe. We theorize that a limited total information transmission capacity in a volume of the universe can lead to the phenomenon of the time dilation, because such a limited information transmission capacity has to be shared by informational contents of all activities occurring in this volume. In other words, in the case that an additional activity occurs in a space volume, the transmission capacity allocated for each activity will reduce, since some portion of the limited capacity has to be used for the additional activities. This will in turn slow down the occurrence of each event in the volume, and an observer living outside of this volume will detect a time dilation inside this volume. Possible activities, which can reduce the transmission capacity allocations and thus lead to time dilation in a space volume, are displacement (shifting) of all content of this volume in space, which is formally referred as the moving frame, and the gravitational effects of massive objects. The term physical activity, in this paper, is used for the all physical events and interactions, which have an informational content.

In progress, we suggest that an unbalance of the transmission capacity distribution inside a volume will result in a force acting inside the volume. For instance, gravity has an effect causing an imbalance in transmission capacity allocations on the space. Because, we know that intensity of gravitational effect reduces with respect to the distance in the space, the reason being that it does not have a homogenous effect inside a space volume. If we assume that conveying informational content of an effect on the network has a correlation with intensity of the effect, specifically, the information transmission capacity allocations for the gravitational effect has to reduce in the space volume as well, with respect to the distance from the mass. This imbalance in distribution of information transmission capacity allocations leads to a time dilation variation inside the volume, and this can result in a gravitational force acting on any object occupying a volume in space. To have a better understanding; let us assume that an electromagnetic wave propagates through a space volume having an imbalance in its distribution of information transmission capacity allocations inside, due to gravitational effects of a nearby object as represented in Figure 1(a). In this case, the time dilation on the side of space, which is closer to sun, becomes larger than the one on the other side of space that is further from sun. Thus, the electromagnetic waves can be observed to move faster in the side of space volume, which is the further from a mass. As a consequence, the bending of light path in the vicinity of objects can occur and this phenomenon becomes equivalent to the gravitational effect of matter on the light path.

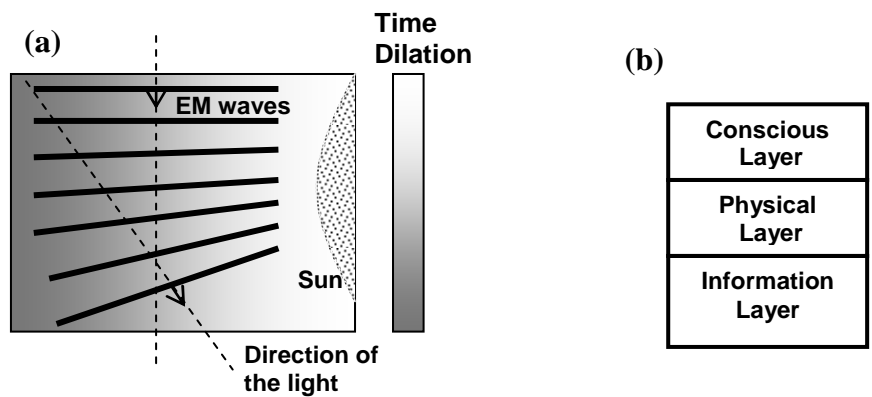

**Figure 1.** (a) Bending of EM wave in the vicinity of Sun. (b) Three layers of Open Universe model.

On the way to better understanding what the universe is, we propose to treat that the problem in a layer-by-layer approach as is done in the Open Systems Interconnection model (OSI model) does [11]. OSI model was developed for the progressive modeling of today's and future communication systems. Such a layer-based perceptive facilitates linking and interacting between different fields of physics from micro to macro or from discrete to continuum. We choose to call such a layer-based modeling an Open Universe model (OU), which is composed of a collection of interconnected layers including various aspects of the universe.

The most studied one of the layers during the science history is the physical layer as it is inside our perception limits. The physical layer contains all the parameters of our observable world (forces, moment, energy, mass, charge…vs) and all physical activities. We suppose that an information layer should be associated with the physical layer and it should provide an informational foundation for the physical layer. The three layered open modeling is represented in Figure 1(b). The conscious layer should rise on the physical layer and this layer has to deal with the issues relating to intelligence, self-awareness or consciousness. In future, optimistically, sub-layers of each layer can be horizontally appended to the table to elaborate the OU model and the additional major layers can be vertically appended to it in order to extend the model coverage.

## 2. Methodology

### 2.1 A Mathematical Foundation for Information Transmission Capacity of the Information Layer

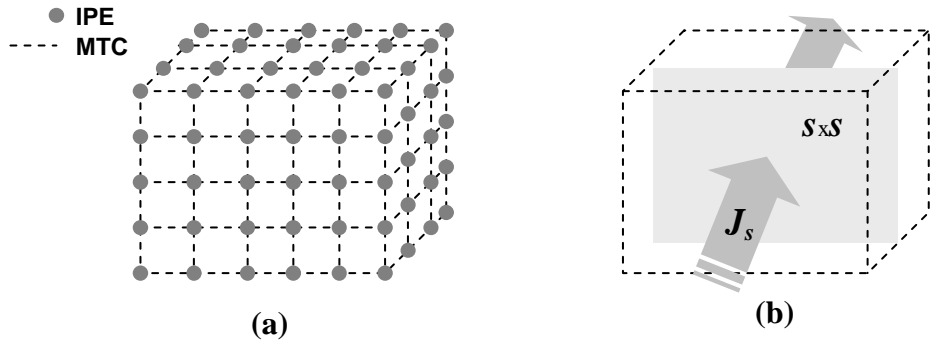

**Figure 2.** (a) A volume of equivalent transmission network. (b) Information fluctuation through a cross section in a volume of space.

An equivalent Transmission Network (ETN) is a distributed network that is assumed to carry and process all informational content of universe. The ETN is composed of (1) Information Processing Elements (IPE) and (2) the Message Transmission Channels (MTC) connecting the each IPE as represented in Figure 2(a). An IPE principally has to simultaneously possess the functions of a transmitter, a receiver, a processor and a router. MTCs are noiseless and finite band-wide transmission links between ETNs.

The ETN can be represented in a variety of scales from complex behavioral models to a Boolean network model. In order to provide synchronization between processing elements and to make the total transmission capacity invariant in all space, a universal clock signal should be used as a trigger for the network operation. In this paper, we try to gain a tangential view on the structure of an ETN, while a more mature and comprehensive description is postponed for a more focused study.

Our basic assumptions for the information layer are as follows:

1. The information layer is contained in the whole space and time, homogeneously. It exhibits the same property in whole space and time.

2. All informational content of the physical universe is assumed to be transmitted via an ETN. In order to comply with Axiom 1, this network is contained by the whole space.

3. Total transmission capacity for a finite volume of universe is limited and invariable in whole space of the physical layer. However, this constant total transmission capacity is sharable by events of physical layer

Let consider a cross section, with the size of $s \times s$, in a volume seen in Figure 2(b). The amount of information transmitted from this cross section in a $\Delta t$ interval of time is denoted with $M_{s \times s}$, in such case, information transmission capacity on the cross section can be written as,

$$J_{s \times s} = \frac{M_{s \times s}}{s^2 \cdot \Delta t}. \tag{1}$$

The parameter $M_{s \times s}$ can be expressed in term of symbols forming the transmitted information (the message) and $J_{s \times s}$ is transmission capacity on the surface $s^2$ in space. The interval of time $\Delta t$ is measured with a universal timer. A universal timer is not exposed to any time dilation, because it runs by a universal clock signal. A conceptual universal timer is a necessity originating from a need of the fastest reference timer, which is used for detecting the time dilation in any frame. For a relative measurement, in practice, an observer timer can be used to relative dilation, but for an absolute measurement, it must be an ideal timer that is free from any time dilation such as the universal timer.

Since the total transmission capacity provided by the network is considered invariable in the whole space according to Axiom 2, one can state that the density of IPE is unvarying and MTCs are identical for all IPEs on the surface $s^2$. Let define $d_b$, the density of IPE , as $d_b = N_b / s^2$, where $N_b$ is the total number of IPE. For such case, the transmission capacity per IPE can be found as,

$$J_b = \frac{J_{s \times s}}{N_b} = \frac{M_{s \times s}}{d \cdot s^4 \cdot \Delta t}. \tag{2}$$

The total transmission capacity per IPE ( $j_b$ ) and for a surface ( $J_{s \times s}$ ) are considered unvarying in all space and time, so they are assumed as a universal constant. But, this total transmission capacity provided by an ETN has to be sharable by all activities in a space volume of the physical world. This sharing of the total transmission capacity by physical activities reduces the allocated capacity per activity and it causes the slowing down performing of the physical activities in the volume. This can be an explanation for the so-called time dilation that was introduced as the slowing clock ticking in frames. A possible transmission capacity sharing policy will be discussed further in the next sections.

## 2.2. An Approach for Capacity Sharing Policy in ETN

Occurrence of the time dilation in the case of the moving frame or the gravity indicates that these activities have a higher priority in transmission capacity allocation process than the internal activity of the frame.

An internal activity for a physical entity defines the set of activities accommodating the frame and continuing. An external activity defines the activities originating from outside of the frame, but which they have an effect on the frame. Let us consider a certain volume in an idle space, which will be considered as being absolutely motionless. Here, the idle space refers the case that there is not any physical activity in this empty space, and correspondingly, it indicates the case where there is not any information fluctuation on ETN. An example for the idle space is the something our physical universe expending through. However, the empty space in our universe is, in fact, an active space, because, there has to be information fluctuation convening interaction of the physical events such as gravitational effects, EM propagation.

Provided that an activity is placed in a certain volume and it retains itself for an adequately long interval of time in this volume, we can say that we had an internal activity in this volume, seen as an absolute motionless frame. Such an internal activity has to fully use the total information transmission capacity of ETN in the volume. If we affect this volume with an external activity such as a gravitational effect, a portion of the total information transmission capacity in the volume will be reserved for this external activity. The rest of transmission capacity can continue to be used by the internal activity. This slows down the performing internal activity in the volume, and results with a time dilation in this volume. According to this scenario, it can be concluded that an external activity should have priority in transmission capacity allocation in the frame.

Let denote the sum of all transmission capacity allocations by $J_{sxs}$. The total transmission capacity $J_{sxs}$ can be analytically expressed as the sum of the transmission capacity allocated for the internal activities and the transmission capacity allocated for external activities:

$$J_{sxs} = J_{\text{int}} + J_{ext}, \qquad (3)$$

where $J_{\text{int}}$ and $J_{ext}$ represent transmission capacity allocations for internal activities and for the effects of the external activities, respectively. If there are more than one activity effective in the volume, the capacity allocations for each effect can be superimposed. So,

$$J_{ext} = \sum_{i=1} J_{ext}^i \text{ and } J_{\text{int}} = \sum_{i=1} J_{\text{int}}^i . \qquad (4)$$

Now, we can reorganize equation (3) as the following:

$$J_{sxs} = J_{\text{int}} + \sum_{i=1} J_{ext}^i , \qquad (5)$$

Movement of frame, which is actually the shifting of all internal activities inside a volume, consumes a portion of the total transmission capacity. Hence, we catalogue it as an external activity and the capacity allocation, denoted by $J_{ext}^1$, will have to be reserved for performing motion of the frame in the space. Others, $J_{ext}^i$ for $i > 1$, are reserved for the gravitational effects of mass and any other effect that has not been discovered, yet. So, the proposed capacity sharing policy can be summarized as listed below:

1. In the case of absolute motionless frames (occupying a certain volume) in an idle space, all total transmission capacity of ETN is allocated for internal activity; that is, $J_{\text{int}} = J_{sxs}$, because $J_{ext} = 0$. In this case, there is not any time dilation in the frame. A timer in this frame can be supposed to be a universal timer.

2. In the case of a moving frame in an idle space, where $J_{ext}^1 > 0$ and $J_{ext}^i = 0$ $(i > 1)$, the additional movement to shift all internal activities in the space consumes a portion of the total transmission capacity and it reduces the capacity allocated for any internal events, so that we can write $J_{\text{int}} = J_{sxs} - J_{ext}^1$. In this case, there occurs a time dilation due to the movement of the frame and therefore, a clock in this frame ticks slower than the clock in the frame described item 1.

3. In the case of an absolute motionless frames (occupying a certain volume) in an active space, let say, surrounded by a mass distribution; the gravitational effects consume some portion of the total transmission capacity of ETN. So, the capacity allocated for the internal activity can be written as $J_{\text{int}} = J_{sxs} + \sum_{i=2} J_{ext}^i$. In this case, there occurs a time dilation due to the gravitational effects, and a clock in this frame ticks slower than the one given by the case in item 1 in this list.

4. In the case of a moving frame in an active space, there can be lots of allocations in the total capacity for serving the external influences, that is, $J_{\text{int}} = J_{sxs} + \sum_{i=1} J_{ext}^i$. Here, $J_{sxs} \geq \sum_{i=1} J_{ext}^i > 0$. The time dilation may be considerably high and the clocks seriously slow down in such frames.

One of the interesting outcomes of this allocation policy is that; an increase in the external activity reduces the transmission capacity reserved for internal activity as represented in Figure 3. In the case of $\sum_{i=1} J_{ext}^i \cong J_{s \times s}$, one obtains $J_{int} \cong 0$, which indicates the case of almost stopping all internal activities in a frame. The physical correspondence of this case can be found in the phenomenon of the prolonging a muon lifetime since its velocity approximates the speed of light [14].

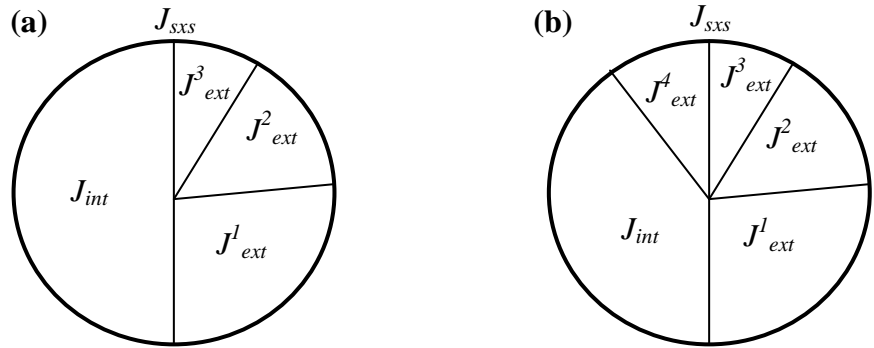

**Figure 3.** Examples of the transmission capacity allocation charts for the proposed sharing policy

## 2.3. An Analysis for Transmission Capacity Allocation of External Activities on the Bases of Relativity Theories

In this section, we will make an analytical treatment of $J_{ext}^i$ on the bases of Relativity Theory [13]. The relativity theory and its modeling for the time dilation have found a wide acceptance in the scientific world.

In the case of a moving frame with a velocity of $v$, the time dilation is given as,

$$\Delta t' = \frac{\Delta t}{\sqrt{1 - \dfrac{v^2}{c^2}}}, \tag{6}$$

Here, we may assume that $\Delta t$ is the time interval for a clock ticking when the frame is in an absolute motionless state in an idle space, and $\Delta t'$ is the time interval for a clock ticking when the frame has a constant velocity of $v$. Both are assumed to be measured by a universal timer. In this frame, let us assume that a clock ticking requires a $M_{s \times s}$ amount of information fluctuating in a surface with the area of $s^2$. Considering equation (6), one can write,

$$\frac{J'_{int}}{J_{int}} = \frac{M_b / (s^2 \Delta t')}{M_b / (s^2 \Delta t)} = \frac{\Delta t}{\Delta t'} = \sqrt{1 - \frac{v^2}{c^2}}, \tag{7}$$

and obtains,

$$J'_{int}(v) = J_{int} \sqrt{1 - \frac{v^2}{c^2}}. \tag{8}$$

Equation (7) yields the $J'_{int}$ transmission capacity reserved for internal activity, when the frame has a constant velocity of $v$. $J_{int}$ is the transmission capacity reserved for internal activity when the frame is in absolute motionless state in an idle space, and it is equal to the total of the transmission capacity $J_{int} = J_{s \times s}$. For this case, equation (8) is reorganized as,

$$J'_{int}(v) = J_{s \times s} \sqrt{1 - \frac{v^2}{c^2}} \ . \tag{9}$$

The capacity allocated for the movement of a frame in idle space, which is denoted by $J^1_{ext}$, can be written as $J^1_{ext} = J_{int} - J'_{int}$. Considering equation (9), we can obtains,

$$J^1_{ext} = (1 - \sqrt{1 - \frac{v^2}{c^2}}) J_{s \times s} . \tag{10}$$

In order to model the gravitational effects on the transmission capacity of ETNs, a formulation derived from the Schwarzchild metric is considered, where the gravitational time dilation in the vicinity of a non-rotating massive spherically-symmetric object is given as

$$\Delta t' = (1 + k / r)^{1/2} \Delta t \ , \tag{11}$$

where $k = 2MG / c^2$ and $r$ is the distance from the center of the spherical mass $M$. For the case of an absolute motionless frame surrounded by an active space that contains a total mass $M$, equation (11) can be reorganized as

$$\frac{J'_{int}}{J_{int}} = \frac{M_b / (s^2 \Delta t')}{M_b / (s^2 \Delta t)} = \frac{\Delta t}{\Delta t'} = \frac{1}{(1 + k_d / r_d)^{1/2}} \ , \tag{12}$$

where, $k_d = \dfrac{2 M_d G}{c^2}$ and $M_d$ is a very small fraction of mass $M$ in the form of sphere and $r_d$ is the distance from a fractional mass $M_d$. From equation (12), one obtains,

$$J'_{int}(r_d, M_d) = J_{int} (\frac{1}{1 + k_d / r_d})^{1/2} \ , \tag{13}$$

where $J'_{int}$ is the transmission capacity reserved for internal activity, when an absolute motionless frame lays on an active space containing a small mass of $M_d$. $J_{int}$ is the transmission capacity reserved for internal activity when the frame is in absolute motionless state in an idle space, and it can be written as $J_{int} = J_{s \times s}$. For this case, equation (13) can be expressed as,

$$J'_{int}(r_d, M_d) = J_{s \times s} (\frac{1}{1 + k_d / r_d})^{1/2}. \tag{14}$$

The mass $M$ can be taken as a sum over all fractional masses $M_d$ in a mass volume $V_M$ and arithmetically expressed as $M = \sum_d M_d$. In such a case, the density of the object of total mass $M$ will be $D_M = M / V_M$. Considering the superposition of gravitational effects, for the mass $M$, $J'_{int}$ on a point of space can be found as,

$$J'_{int}(M) = J_{s \times s} \sum_d (\frac{1}{1 + k_d / r_d})^{1/2}. \tag{15}$$

The capacity allocated for gravitational field at distance $r$ from a spherical mass $M^i$ can be written as $J^i_{ext} = J_{int} - J'_{int}$, $i > 1$. Considering equation (15), we can obtains,

$$J^i_{ext}(M^i) = (1 - \sum_d (\frac{1}{1 + k^i_d / r^i_d})^{1/2}) J_{s \times s} . \tag{16}$$

In order to obtain a general solution for $J_{int}$, we can reorganize equation (5) as the following.

$$J_{\text{int}} = J_{s \times s} - \sum_{i=1} J_{ext}^i \qquad (17)$$

As an example, we can consider the case that a frame is moving at a velocity $v$ in an active space containing a total mass of $M$, in which case, $J_{\text{int}}$ capacity can be found as,

$$J_{\text{int}} = (\sqrt{1 - \frac{v^2}{c^2}} + \sum_d (\frac{1}{1 + k_d / r_d})^{1/2} - 1) J_{s \times s}. \qquad (18)$$

When $L$ number of objects with the mass $M^i$ reside around the frame, $J_{\text{int}}$ can be write as

$$J_{\text{int}} = (\sqrt{1 - \frac{v^2}{c^2}} + \sum_{i=2}^{L+1} \sum_d (\frac{1}{1 + k_d^i / r_d^i})^{1/2} - L) J_{s \times s}, \qquad (19)$$

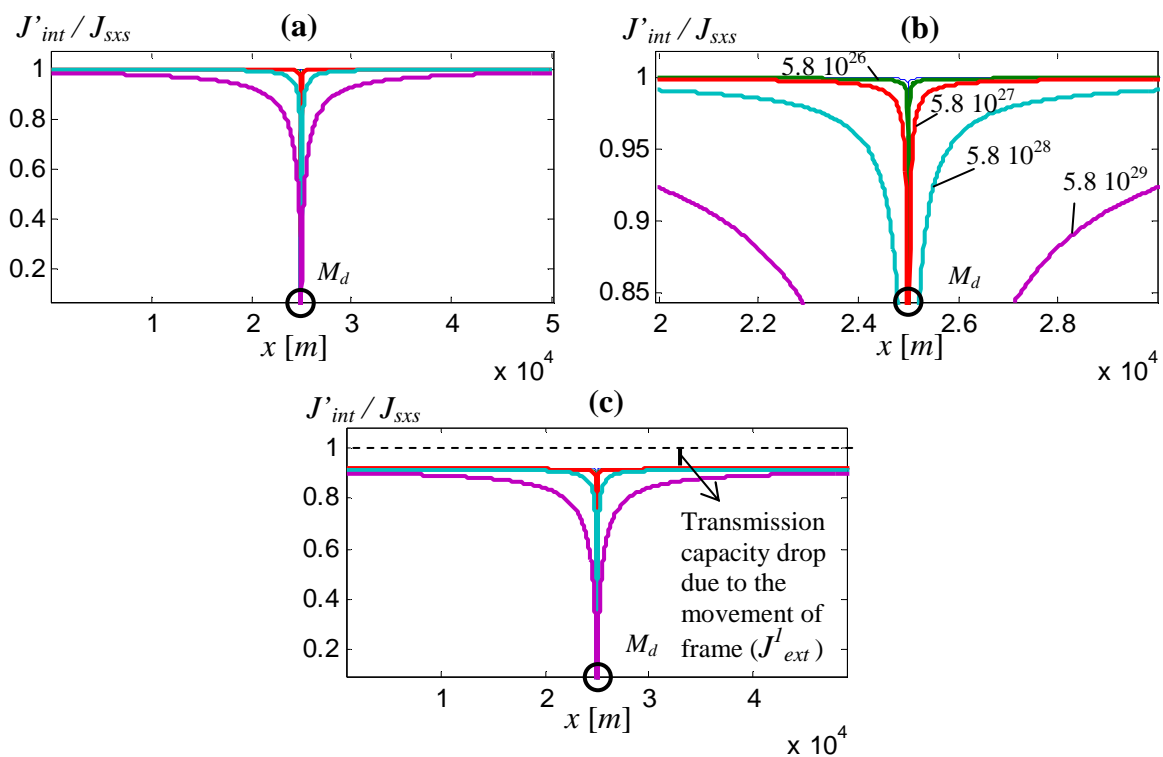

**Figure 4.** Normalized transmission capacity allocation ($J'_{\text{int}} / J_{s \times s}$) is drawn for the various mass placed on one-dimensional space $x$ in (a) and its zoomed version given in (b). In the case of the moving frame at velocity of $0.4c$, the same curvatures is obtained with a drop of $J_{ext}^1$ from one in (c).

## 2.4. A Discussion of the Gravitational Force Appearing In the Physical Layer

The matters or electromagnetic waves sustain as an entity in the spacetime. This is why; informational content accompanying them can be considered to convey on ETN. We suggest that when the space volumes of these sustaining entities have an unbalanced $J'_{\text{int}}$ distribution, this can cause the appearance of force in physical layer.

For the sake of simplicity, we provide an explanation for the bending of light nearby a huge mass based on our set of axioms. As in Figure 1(a), $J_{\text{int}}$ decreases near the sun and, in physical layer, it leads to an increase of the effective time dilation near the sun than further from it. Since, the electromagnetic wave forming the light can move more slowly near the sun than the away from it, the light follows a curving path around the sun, which we interpreter it as a gravitational force attracting the light through the sun.

Let us consider a space where a mass $dm$ is at rest, and compare $J'_{int}$ at distance of $r_1$ and $r_2$ from the mass center and resulting time dilations $\Delta t'(r_1)$ and $\Delta t'(r_2)$, respectively. For a couple of identical physical activations residing at distance of $r_1$ and $r_2$, the rate of the speed of activation can be written as

$$\frac{J'_{int}(r_1)}{J'_{int}(r_2)} = \frac{\Delta t'(r_2)}{\Delta t'(r_1)} = \frac{V(r_1)}{V(r_2)} = (\frac{1 + k_d / r_2}{1 + k_d / r_1})^{1/2}, \tag{20}$$

where $V(r_1) = x / \Delta t'(r_1)$ and $V(r_2) = x / \Delta t'(r_2)$, in its simplistic form, is speed of identical physical activities, respectively. $x$ represents the any measurable physical quality in this activity. In order to make a comparison with the case of an idle space, we can write following equation,

$$\frac{J'_{int}(r_1)}{J_{s \times s}} = \frac{\Delta t'(r_2)}{\Delta t} = \frac{V(r_1)}{V_0}, \tag{21}$$

where $V_0$ is the velocity of an identical activation in an idle space. Using equation (14), we can write the following,

$$\frac{V(r_1)}{V_0} = \frac{1}{(1 + k_d / r_1)^{1/2}}. \tag{22}$$

Therefore, according to the proposed model, the time dilation depends on the transmission capacity allocation and when distribution of the transmission capacity allocation in the whole space fails to be uniform, we can state that the speed of light varies in the space. However, if the frame of observer and the frame of the light have the same capacity allocation or if the observer and the light beam are in the same frame, the observer measures the speed of light as the value of the constant $c$, which is agreed as the speed of light in the vacuum. But, any difference in allocated internal capacities between the observer frame and the frame, where the light is traveling, will lead the speed of the observed light to be different from the value of the constant $c$. Today, we know that the light can be slowed down, experimentally. We want to notice that if the speed of light in space is limited, it will possibly be the result of the limits of transmission capacity allocation.

Full discussion of the gravitational effects is more challenging and may need additional assumptions, in view of the fact that the matter is much more sophisticated in structure than the light. Since all information fluctuation performed for sustaining matter in space-time is mainly confined to the volume occupied by the matter, such retaining fluctuations must follow periodical paths representing an information circulation in the volume occupied by matter.

The most idealized path formations in the volume of matter are circular paths, or ellipsoidal paths, however it may be possible to find more irregular and complicated information fluctuation paths on EIT. The point we which to stress here is that it is the imbalance of the $J_{int}$ distribution in the volume occupied by the matter, that makes some physical events to occur faster than the other part due to unbalanced time dilation distribution in the volume as shown by the analytically derived equation (20). This can result in an unbalanced state the overall motion of matter such that the side faced more time dilation moves slower than the other sides and this can change the direction of the matter through the slower side in a manner similar with that discussed for the light bending near the sun. Figure 5 illustrates such effects for the both light and matter.

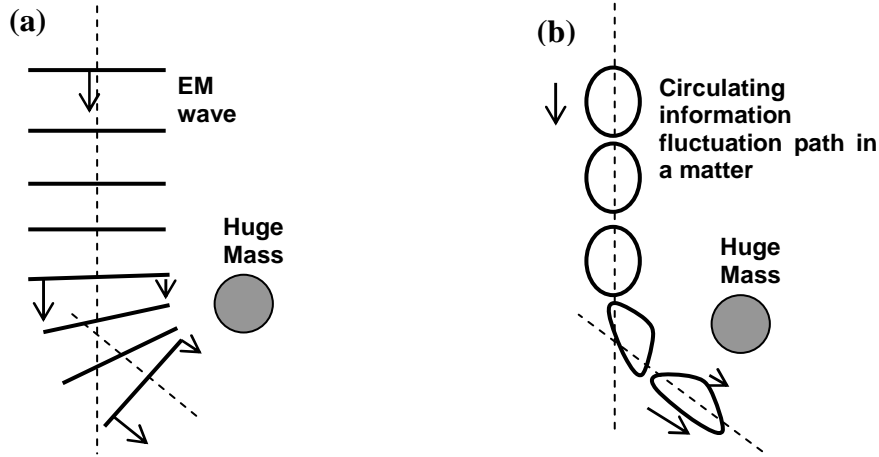

**Figure 5.** (a) Attraction of Electromagnetic (EM) waves by a mass. (b) Attraction of a moving matter, as a result of the imbalanced distribution of transmission capacity on the circular information fluctuation path on ETN.

In order to derive the gravitational force acting on a body with a mass $m$ from the well known equation, $F = G\dfrac{m \cdot M}{r^2}$, one should consider the mass of very small volumes in the mass $m$, which is denoted by $dm$. In such a small volumes of $m$, deviation of $J_{\text{int}}$ inside the small volumes can be ignorable and the volumes can be assumed to have an average $J_{\text{int}}$. Using equation (14) and $\dfrac{dF}{dm}r = G\dfrac{M}{r}$, can obtain the force $F$ analytically as follows,

$$\frac{dF}{dm} = \frac{c^2}{2r}\left(\frac{J_{sxs}}{J'_{\text{int}}} - 1\right),\tag{23}$$

In this equation, the term $\left(\dfrac{J_{sxs}}{J'_{\text{int}}} - 1\right)$ is always grater than zero and it stands for the gravitational of effects of the mass $M$.

The relation of $dE = dm \cdot c^2$, energy of mass, is applied to equation (23), we can see that the gravitational force can be expressed depending on mass energy.

$$dF = \frac{dE}{2r}\left(\frac{J_{sxs}}{J'_{\text{int}}} - 1\right),\tag{24}$$

By integrating equation (23) ,we obtains the total gravitational force on $m$ in the form,

$$F = \frac{c^2}{2}\int_0^m \left(\frac{J_{sxs}}{J'_{\text{int}}} - 1\right)\frac{1}{r}dm.\tag{25}$$

### 2.5. A Virtual Experiment with FDTD Wave Simulation

In Figure 6, we present a FDTD simulation result from wave propagation on a medium of particles under effect of gravitational time dilation. Wave on such a medium, made of a bulk of free particles, is characterized by

$$\alpha \cdot \frac{\partial \vec{\xi}}{\partial t} = -\nabla \psi \ \text{ and } \ \frac{\partial \psi}{\partial t} = -\beta \cdot \nabla \cdot \vec{\xi} , \tag{26}$$

where the parameters $\alpha$ and $\beta$ specify medium properties relating to density of particles and resistance of particles against the compression. $\vec{\xi}$ denotes particle velocity vectors and $\psi$ is a scalar field denoting compression rate of particle in space. The compression waves on this homogenous medium propagates at a constant velocity that is determined by $V_s = \sqrt{\beta / \alpha}$. The system defined by equation (26) was used in the simulation of acoustic waves.

$\vec{\xi}$ vectors can be considered to have a link with information fluctuation on ETN of information layer and the $\psi$ scalar can be considered as a physical state that can be observable in the physical layer. Thereby, the space is assumed to be a bulk of free particles and the physically observable activity is assumed to be a wave-like motif of particle compression fields. In Figure 6(f), a grid structure of this system for a discrete FDTD simulation is presented.

Wave velocity at a location of medium, $V_s(r)$, has been written with respect to equation (22) as $V_s'(r) = V_s \cdot (J_{int}'(r) / J_{sxs})$, which implies that the variation in transmission capacity sharing on ETN alters the speed of activities on the medium.

This virtual experiment setup provides us a simulation of the effects of an unbalanced transmission capacity allocation on the physical activity, which is the particle wave propagation in our case. To this purpose, a $M_d$ mass, exhibiting adequately strong gravitational effect, was placed in the medium of free particles. In the wave propagation simulation done for a two-dimensional plane of the medium, it was seen from the results given in Figure 6(a) and (c) that the waves were bended around the point, where the $M_d$ mass resides. This can be interpreted in the physical layer as a gravitational force attracting the waves around the $M_d$ mass or as a space-time curvature on the path of the wave in the vicinity of mass. But, in the information layer, it can be seen as an imbalance on the transmission capacity allocation on the medium. It leads to an alteration in the speed of physical activity. Accordingly, we can observe in the simulation results that the wavelength is diminishing around the mass. It indicates a higher time dilation in the vicinity of mass.

Normalized distribution of the transmission capacity allocations on the medium, which is the rate of $J_{int}'(r) / J_{sxs}$, is given in Figure 6(b) and 6(d). In Figure 6(e), we present the wave of particles, for the case that there is not any adequately large mass on the medium. Here, $J_{int}' / J_{sxs} \cong 1$ holds everywhere in the homogeneous medium, and the attraction effect of gravity on the wave was seen to disappear in this simulation.

We hope to do a more focused study addressing the particle system defined by equation (26) in the future. Products of such a study may extend our understanding on the waves-particles interactions in fluid like systems.

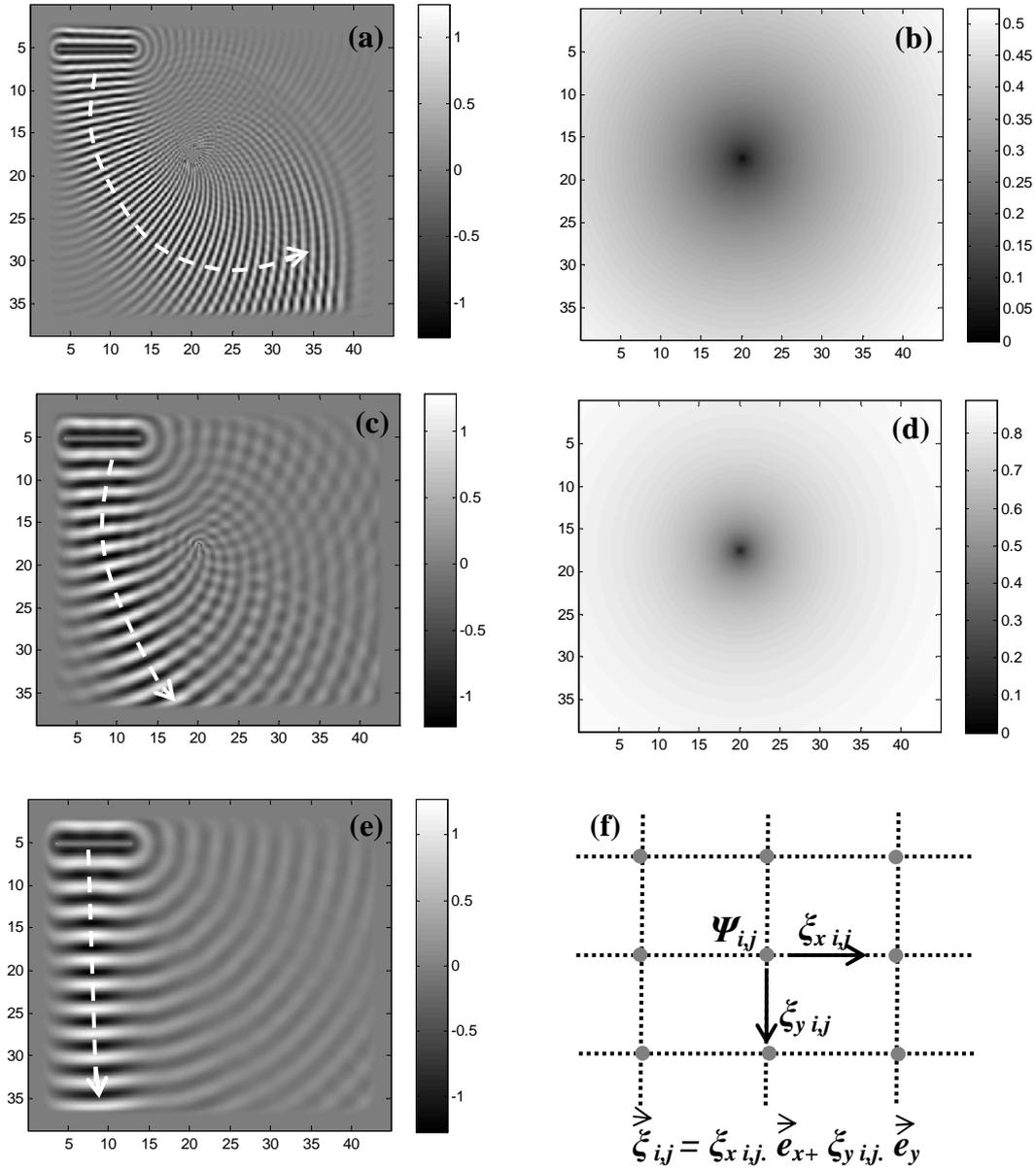

**Figure 6.** In (a) and (c), FDTD simulation results showing wave bending due to gravitational effect are presented and In (b) and (d), corresponding distribution of the normalized transmission capacity allocations ($J'_{\text{int}} / J_{sxs}$) for gravitational effect are presented, respectively. (For (b) $M_d = 5.8\ 10^{29}$ kg for (d) $M_d = 5.8\ 10^{28}$ kg ) The image (e) illustrates the wave propagation for $J'_{\text{int}} / J_{sxs} \cong 1$, which infers that there is not a strong imbalanced gravitational effect on the medium. (f) illustrates grid of FDTD method. ($\vec{e}_x$ and $\vec{e}_y$ are base vector on the plane)

## 2.6. A Discussion on IPE Density on a Planck Area

Scale of entities in universe hierarchically varies. For example, the sub-atomic particles make up atoms, atoms form the matter and planets are composed of matter. This is why, size of frames can be taken into account depending on what we want to observe. If an atom is the observable, frame can be taken to be a volume of space on the size of an atom. In this case, the inner activities of the frame will be the set of all activities maintaining the atom. If a planet is observed, the frame has to be in the size of planet.

Let a frame or a volume be continuously sub-divided into sub-frames or sub-volumes. The volume of these sub-frames can be expressed as $\frac{s^3}{2^q}$, where $q$ is the number of sub-divisions. In such a case, we can observe that the internal activity of each sub-frames varies and, off course, $J_{int}$ of each sub-frame will show a volatility due to mutual interactions between sub-frames and it will show a discrepancy from one to the other. It indicates that the time dilation distribution in a volume is not uniformly distributed, and, it can even be said to be time-varying, because of the influence of volatility in all neighbor sub-frames around it. In this point of view, we can state that if we continue splitting sub-frames, we can expect to see that the volatility will increase in smaller sub-frames. So, increasing the size of frame as a superposition of state of more sub-frames makes an averaging effect on volatility in $J_{int}$ and it results in gain the frame a more stable appearance in statistically. In macro scale, gravitational field is dependent of distance ($r$); it becomes a substantial factor for inhomogeneous time dilation distribution in a frame. But, we can see that $J_{ext}^1$ provides a constant affect because it has the same affection on all child-frames of a mother frame, therefore it is independent of distance.

After seeing the frame can be scalable, a question comes to minds: How much can we diminish a frame or a volume? The answer in the ETN model point of view, it can diminish until a frame contains an IPE. On the other side, from the physical layer point of view, the Planck length can define a limit of minimal sizes in universe. In order that a Planck volume in universe can be detectable or be observable, it should be adequately large to involve in a physically meaningful activity, which means to process and convey an informational content about the activity. So, from the ETN point of view, a Planck area should have at least one IPE, and accordingly the IPE density should the satisfy following condition,

$$d_b \geq 1/s_p{}^2, \tag{27}$$

where $s_p{}^2$ represents a Planck area. Larger $d_b$ implies placing much more IPEs into a Planck volume.

## 3. A Few Experiments for the Information Layer:

Measurement of the time dilation in a laboratory environment has some complications. The major one is to generate a measurable time difference between a reference system and a test system in a laboratory. Main reason for that, the size of laboratory environment on the earth can not be enough large to move a clock mechanism at the velocity that result with a measurable time difference between a reference clock mechanism and a moving test clock system. At this point, there needs for the technique that is highly responsive to the very small time dilations, and yet the experiment can be conduct in the limited space of laboratory environment. In order to manage these objectives, an optical experimental setup is proposed, which is, in fact, an adopted version of well known Young's double split experiment. In this interferometrical method, we aim to detect even small variation in the phase of light wave by observing interference patterns of a reference wave and a test wave. The point here is that, if the path of test waves exhibits an additional in time dilatation, this will cause a phase shift on test waves, and such a phase shifts on test waves results with a change in interference pattern of the reference and the test waves. Thus, amount of the time dilation on the test path can be estimated by considering the change in traces of constructive or destructive interference on the interference pattern.

We introduce two experimental setups for the estimation time dilation and the finding signs of ETN. A direct and interactive observation of ETN have some difficulties: firstly, the experimental system is also made of ETN, so any attempt to access a fraction of ETN has an influence on content of this ETN fraction, such as information leakages. Moreover, we can not today effectively isolate a fraction of ETN from the rest of it to do direct test on it. Inevitable information leakages trough the tested ETN fraction, which may come from the universe and our measurement systems, will bring an uncertainty that drift us to the worlds of probabilistic models and to the problem of dealing with highly noisy behavior.

### 3.1. Measurement of Variation in Speed of Waves:

In order to measure change in wave velocity, the double split Young experiment is benefited as given in Figure 7.

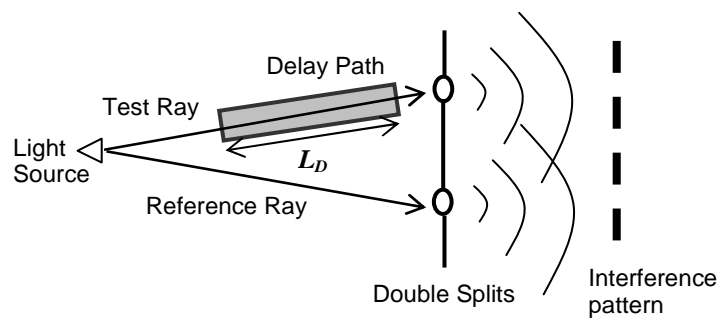

**Figure 7.** Experimental setup for the measurement of light velocity variation for the estimation of the time dilation along the delay path

For a chosen $X$ point on the interference pattern, the path difference, $\Delta L$, between test ray and reference ray from the source to the point $X$ is written as $\Delta L = L_t - L_r$. Here, $L_t$ is length of test ray from source to the point $X$ and $L_r$ is length of reference ray from source to the point $X$. If a constructive interference falls on the point $X$, $\Delta L$ was known to be as,

$$\Delta L = m\lambda, \tag{28}$$

,where $\lambda$ is wavelength and $m$ is an integer number. If a destructive interference falls on the point $X$, $\Delta L$ was known to be as,

$$\Delta L = m\frac{\lambda}{2}, \tag{29}$$

Considering equation (28) and (29), we can easily state that if $X$ is on a constructive interference, the phase shift, denoted by $\Delta\varphi$, between test and reference waves on the point $X$ is zero and if $X$ is on a destructive interference, the phase shift between test wave and reference waves on the point $X$ is $\pi$. After reminding these basics, let us derive expression that can yield wave velocity differentiation on the paths of test and reference rays. This expression can used to estimate the relative time dilation of a test path according to a reference path so that the wave velocity on the test path has to vary with respect to the time dilation on the delay path.

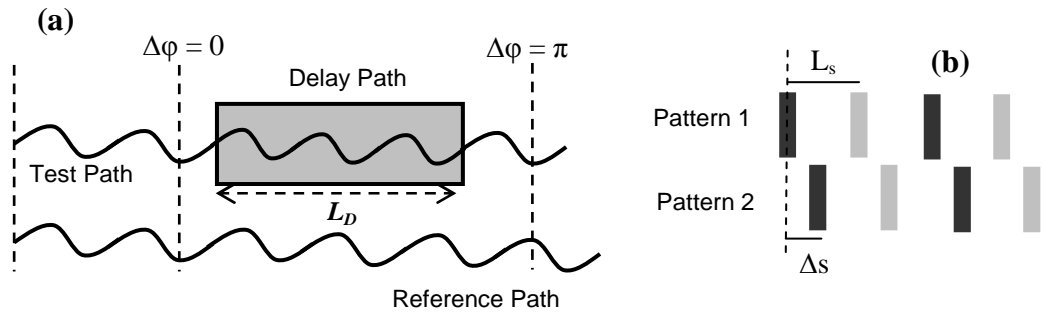

**Figure 8.** (a) Phase shift of wave by means of a delay path. (b) $\Delta s$ shift in interference pattern

Let assume a delay path produce a $\Delta\varphi = \pi$ phase delay in the waves as seen in Figure 8. It results shifting the constructive interferences with the destructive interferences on the interference patterns. For the wave velocity in the delay path, one can write the following,

$$V_D = (1 - \frac{\lambda}{2L_D}) \cdot V_r, \tag{30}$$

where $V_D$ is wave velocity in the delay path and $V_r$ is wave velocity in the reference path. $L_D$ is the length of delay path and it should be larger than several $\lambda$.

For the shifting less than the distance between a constructive and a destructive interferences, wave velocity can be reorganized as

$$V_D = (1 - \frac{\gamma\lambda}{2L_D}) \cdot V_r. \tag{31}$$

$\gamma$ is the rate of the pattern shifting in a partial distance between a consecutive constructive interference and a destructive interference and it can be expressed as $\gamma = \Delta s / L_s$ under condition of $\Delta s \leq L_s$. $L_s$ denotes the distance between a constructive interference and a destructive interference on the interference pattern as shown in Figure 8(b) and $\Delta s$ length of shifting of a constructive interference or a destructive interference.

In its simplistic and conventional form, the speed of an physical activities was accounted as $V = x/\Delta t$. $x$ represents the any measurable physical quality in this activity, and in our case, we suppose it as the traveling distance in term of ETN elements. $\Delta t$ is time interval in term of an universal timer. If the traveling of a light on the reference and test paths is viewed to be the physical activities, speed of these activities, which are the velocity of the light on the paths, can be written for the reference path as $V_r = x_r/\Delta t_r$ and for the test path as $V_t = x_t/\Delta t_t$. The ETN element is assumed to be identical on the whole space according to the item 1. This means that the reference path and test path are identical in term of ETN elements, so we can write $x_r = x_t$, and in that case, the rate of light velocities on the paths can be easily obtained as,

$$\frac{V_t}{V_r} = \frac{\Delta t_r}{\Delta t_t}. \tag{32}$$

In the experiment, a delay path causes a time dilation difference, so we can consider delay path as the test path. By doing so and reorganizing equation (31) by (32), a roughly estimation of the relative time dilation on the delay path, which is normalized to the reference path, can be found as:

$$\frac{\Delta t_t}{\Delta t_r} = \frac{1}{1 - \dfrac{\gamma \lambda}{2L_D}}. \tag{33}$$

In Table 1, relative time dilation estimations from the equation (33) are listed with respect to several $\gamma$.

**Table 1.** The relative time dilation and for $L_D = 100\lambda$ and various $\gamma$.

| Relative time dilation ($\Delta t_t / \Delta t_r$) | Rate of wave velocity ($V_D / V_r$) | Rate of the shifting ($\gamma$) |
|---|---|---|
| 1.001001 | 0.9990 | 0.2 |
| 1.002506 | 0.9975 | 0.5 |
| 1.003512 | 0.9965 | 0.7 |
| 1.005025 | 0.9950 | 1.0 |

In the following section, two experiments for two different delay path configurations are discussed. In the first experiment, a delay path is formed by a rotating ring of test path, in which, we plan to observe the time dilation of a moving frame. The second one is planed to see if the EM interference can result with a time dilation. It is expected to occur due to the additional capacity allocations for EM interference on delay path. If such a time dilation resulting from EM interference can be detected, it will be an evident sign of the transmission capacity sharing of ETN.

## 3.2. An Experiment Setup for Estimating Time Dilation for Moving Systems

In this experiment, shown in Figure 9, a light from a source, which is placed the center of a rotating wheel, split into two path: The shortest path is length of $r_1$ and it is the reference path to double split. The longest path is approximately length of $n \cdot 2\pi r_2 + 2r_2$ and it forms the test path. Here, $n$ is the number of roll around the wheel. To manage $n > 1$, fiber optic wire can be used.

Main objective in this experiment is to form moving frames having different time dilation. For this proposes, a rotating wheel is utilized to obtain difference in linear velocity of a reference path and a test path. Provided that wheel rotates with an angular velocity $w$, the rate of linear velocities with respect to radius of circular path on the wheel can be simply written as,

$$\frac{v_1}{v_2} = \frac{r_1}{r_2} \, . \tag{34}$$

The slowest frame, which is reference path, can be obtained by using the path with $r_1$ and the faster frame, which is the test path, can be obtained by using the path with several roll at $r_2$ as seen in Figure 9. Thus, we have two frame with different speed in a area of $\pi r_2^{\,2}$.

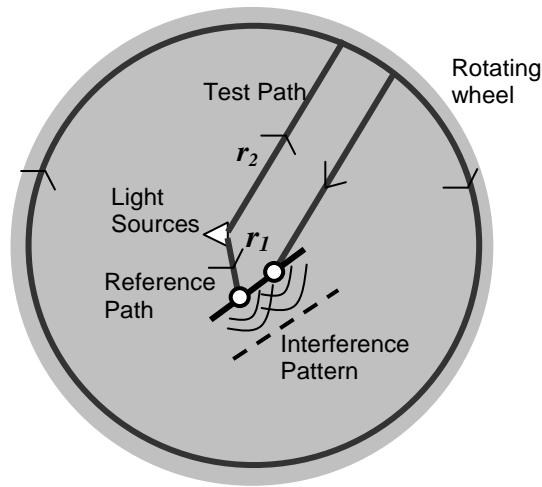

**Figure 9.** Experimental setup for the estimation of the time dilation from special relativity theory

In experiment, firstly, a stable interference pattern, called pattern 1, is obtained while the wheel is still and than, another stable interference pattern, called pattern 2, is obtained while the wheel is rotating at an adequately large velocity. Considering the shifting in interference patterns and Equation (33), a time dilation estimation can roughly be given as,

$$\frac{\Delta t_t}{\Delta t_r} = \frac{1}{1 - \dfrac{\gamma \lambda}{2(n \cdot 2\pi r_2 + 2r_2 - r_1)}} \, . \tag{35}$$

Here, the length of delay path was taken as the all the way of test path, that is, approximately $L_D \cong n \cdot 2\pi r_2 + 2r_2 - r_1$.

Let figure out how much shifting that we can obtains in the interference pattern, when a experiment configured with the parameters as $r_1 = 0.1$ m, $r_2 = 1$ m, $w = 11850$ rad/sn (113159 rpm), $\lambda = 100$ nm, $n = 10$. With consideration of time dilation of special relativity, one can theoretically calculate the rate of time dilation between test and reference path by using the following equation,

$$\frac{\Delta t_t}{\Delta t_r} = \sqrt{\frac{c^2 - v_1{}^2}{c^2 - v_2{}^2}} \,, \tag{36}$$

as $\dfrac{\Delta t_t}{\Delta t_r} = \sqrt{\dfrac{(3.10^8)^2 - (1185)^2}{(3.10^8)^2 - (11850)^2}} \cong 1.0000000007723237$ . When this time dilation is applied to

equation (35), $\gamma \cong 0.99987 \cong 1$ is obtained. This infers that when the wheel rotated at $w = 11850$ rad/sn, constructive and destructive patterns are expected to be substituted in the interference pattern. In the case, such a phase shift in interference patter is obtained, a time dilation resulting from velocity of the frame can be said to be observed in the laboratory. Besides, if obtained phase shift is not consistent with this theoretical calculation, this inconsistency may let us to researching new factors effecting time dilation. Perhaps, the force appearing owing to sharply rotational motion can have an unknown effects on time dilation.

### 3.3. An Experiment Setup for Investigating Effect of EM Interference on Time Dilation

The proposed model tells us that all physical activity using an ETN has to consume much or less a transmission capacity of the ETN and this will lead much or less a time dilation in physical layer. For instance, in the case that two EM waves occupies the same volume in space. The model implies that the intersection of two EM waves should affect the speed of the light at the intersection volume due to the sharing limited transmission capacity. In literature, we haven't seen any report claiming such an effect of EM waves. Maybe, its reason was that the resulting time dilation in intersection volume is so small that we can not detect it, yet. Nevertheless, experimental validation of a reduction in speed of the light as a result of an EM interference will be a millstone for the proposed model. In Figure 10, we suggest an experiment that is benefit from the technique described in Figure 7. The EM interference chamber is considered as the delay path when it is active. The EM interference chamber should be designed to provide a strong EM interference at desired frequencies and it should be shielded to prevent possible EM leakages from outside of chamber.

The objective in this experiment is to test if the enabling or disabling the EM interference in the vacuum chamber results with a change in the interference pattern. If a change in the pattern is observed, it will be indication of that the speed of the light is changed by an EM interference, so it can be viewed to be a sign for the transmission capacity allocation for the EM waves.

We encourage the researchers working on field of experimental optics to test and to report the result of this interesting experiment. If EM waves interference cause a time dilation, and likely, if the time dilation originating from EM interference is adequately large to cause a detectable phase shift on the light waves passing through the chamber, the interference pattern appearing on screen will be seen to change by the enabling/disabling of EM interference in the chamber.

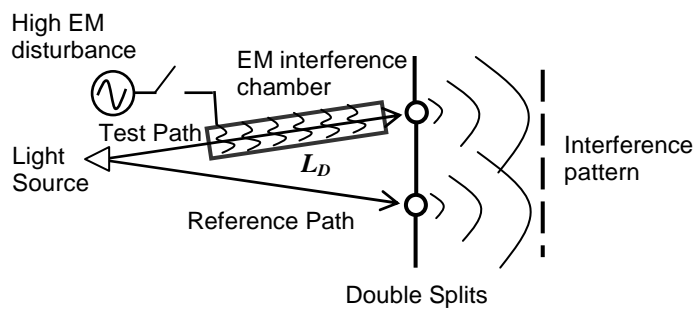

**Figure 10**. Experimental setup for investigating the time dilation from EM interference

## 4. Discussions and Conclusions

In this study, we investigated effects of a limited information transmission capacity and we demonstrated connections of the time dilation and the gravitational force with the limited information transmission capacity.

This study is a theoretical attempt to figure out features of the network that is capable of conveying the informational content that is required for the construction of our physical world. We present a distributed information processing network accessing the whole space to convey the informational contents of physical events over space-time, and demonstrate that sharing the limited information transmission capacity can be responsible from the time dilation in the physical world. A simple transmission capacity sharing policy was discussed for ETN and its analytical foundation, considering effects of movement of the frame and effects of the gravitation, is derived on the bases of Relativity theory.

We believe that open universe modeling can put a good perceptive to facilitate dealing with the complications faced in the research for a Theory of Everything. Modeling the universe as the combination of associated layers may ease inter-disciplinary studies on the subject. The presented perspective in this study, which attempts to connect the information layer (ETN) to the physical layer (Physical world), can contribute to further researches on a possible universal cellular automaton.

The suggested model can give reasonable explanation on some phenomenon relating to space-time; however, we accept that there is still a need for more exclusive and strong evidences that directly indicate the existence of an ETN. For the future study, we suggest a few experimental setups that may help uncover some features of the ETN. We believe that more mature and accurate capacity sharing policies can be developed in with the light of future experimental findings.

Our discussion on the subject is in conceptual level and it needs more effort to reach it a scientific maturity.

**Acknowledgment**

This study was done for the memory of Dr. Serdar Onur ALAGOZ in OncuBilim Algorithm and Systems Labs. We thank Raptis Theofanis for his corrections, which greatly improve readability of the paper.